\documentclass[aps,prd,preprint,superscriptaddress,tightenlines,nofootinbib]{revtex4}
\usepackage{epsfig}
\usepackage{graphicx}
\usepackage{dcolumn}
\usepackage{bm}

\newcommand{\etal}{{\it et al.}}

\begin{document}

\preprint{CLEO CONF 04-11}   
\preprint{ICHEP04 ABS11-0776}      

\title{\LARGE Measurement of ${\cal{B}}(D^+\to\mu^+\nu)$ and the
Pseudoscalar Decay Constant $f_{D^+}$}
\thanks{Submitted to the 32$^{\rm nd}$ International Conference on High Energy Physics, Aug 2004, Beijing}
\author{D.~Besson}
\affiliation{University of Kansas, Lawrence, Kansas 66045}
\author{K.~Y.~Gao}
\author{D.~T.~Gong}
\author{Y.~Kubota}
\author{B.W.~Lang}
\author{S.~Z.~Li}
\author{R.~Poling}
\author{A.~W.~Scott}
\author{A.~Smith}
\author{C.~J.~Stepaniak}
\author{J.~Urheim}
\affiliation{University of Minnesota, Minneapolis, Minnesota
55455}
% ...
\collaboration{CLEO Collaboration} 
\noaffiliation

\date{\today}

\begin{abstract}
In 60 pb$^{-1}$ of data taken on the $\psi(3770)$ resonance with
the CLEO-c detector, we find 8 $D^+\to\mu^+\nu$ event candidates
that are mostly signal, containing only 1 estimated background.
Using this statistically compelling sample, we measure preliminary
values of ${\cal{B}}(D^+\to\mu^+\nu)=(3.5\pm 1.4 \pm 0.6)\times
10^{-4}$, and determine $f_{D^+}=(201\pm 41\pm 17)$ MeV.
\end{abstract}

\maketitle
\section{Introduction}
Measuring purely leptonic decays of heavy mesons allows the
determination of  meson decay constants, which connect measured quantities,
such as the $B\bar{B}$  mixing ratio, to CKM matrix elements.
Currently, it is not possible to determine $f_B$ experimentally from
leptonic $B$ decays, so theoretical calculations of $f_B$ must be used.
The most promising of these calculations involves lattice QCD \cite{Davies} \cite{Lattice:Milc} \cite{Lattice:UKQCD}, though there are other methods \cite{Equations} \cite{Chiral} \cite{Sumrules} \cite{Quarkmodel} \cite{Isospin}.

Measurements of pseudoscalar decay
constants such as $f_{D^+}$ provide checks on these calculations and
help discriminate among different models.

\begin{figure}[htbp]
\centerline{ \epsfxsize=3.0in \epsffile{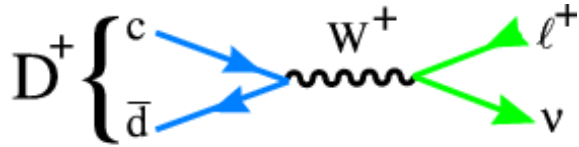} }
\caption{The decay diagram for $D^+\to \mu^+\nu$.} \label{Dptomunu}
\end{figure}

The decay diagram for $D^+\to \mu^+\nu$ is shown in
Fig.~\ref{Dptomunu}. The decay rate is given by \cite{Formula1}
\begin{equation}
\Gamma(D^+\to l^+\nu) = {{G_F^2}\over 8\pi}f_{D^+}^2m_l^2M_{D^+}
\left(1-{m_l^2\over M_{D^+}^2}\right)^2 \left|V_{cd}\right|^2~~~,
\label{eq:equ_rate}
\end{equation}
where $M_{D^+}$ is the $D^+$ mass, $m_l$ is the mass of the final
state lepton, $V_{cd}$ is a CKM matrix element equal to 0.2205
\cite{PDG}, and $G_F$ is the Fermi coupling constant. Various
theoretical predictions of  $f_{D^+}$ range from 190 MeV to 350
MeV. Because of helicity suppression, the electron mode $D^+ \to
e^+\nu$ has a very small rate. The relative widths are
$3.2:1:2.4\times 10^{-5}$ for the $\tau^+ \nu$, $\mu^+ \nu$ and
$e^+ \nu$ final states, respectively. Unfortunately the mode with
the largest branching fraction,
 $\tau^+\nu$, has at least
 two neutrinos in the final state and is difficult to detect.

\section{The CLEO-c Detector}

The CLEO-c detector is equipped to measure the momenta and
direction of charged particles, identify charged hadrons, detect
photons and measure with good precision their directions and
energies. Muons above 1.1 GeV can also be identified. The detector
is almost cylindrically symmetric with everything but the muon
detector inside a superconducting magnet coil run at a current
that produces an almost uniform 1.0 T field. The detector consists
of a six-layer wire drift chamber at small radius that is low
mass, suitable for these relatively low energies. It is followed
by a 47-layer drift chamber; both chambers use a gas mixture of
60\% Helium and 40\% Propane. These two devices measure charged
track three-momenta with excellent accuracy. The drift chamber
also measures energy loss, dE/dx, that is used to identify charged
tracks below about 0.7 GeV/c. After the drift chamber there is a
Ring Imaging Cherenkov Detector (RICH) \cite{RICH} , that
identifies charged particles over most of their momentum range.
The RICH is surrounded by a Thallium doped CsI crystal array
consisting of about 8000 tapered crystals 30 cm long and about 5x5
cm$^2$ at the rear.

\section{Data Sample and Signal Selection}

In this study we use 60 pb$^{-1}$ of CLEO-c data recorded at the
$\psi''$ resonance (3.770 GeV). These events consist of a mixture
of $D^+D^-$, $D^o\overline{D}^o$ and continuum events. There also
may be small amounts of $\tau^+\tau^-$ pairs and two-photon
events.

We examine all the recorded events and retain
those containing at least one charged $D$ candidate in the modes listed in
Table~\ref{tab:Drecon}. The selection criteria are described in detail in what
follows.


All acceptable track candidates must have a helical trajectory that
approaches the event origin within a distance of 0.005 m in the
azimuthal projection and 0.05 m in the polar projection, where the
azimuthal projection is in the bend view of the solenoidal magnet.
Each track must posses at least 50\% of the hits expected to be on a
track and its polar direction must have cosine with respect to the beam direction of magnitude $<0.93$.


We use both charged particle ionization loss in the drift chamber
(dE/dx) and RICH information to identify kaons and pions used to
fully reconstruct $D$ mesons. The RICH is used for momenta larger
than 0.55 GeV. Information on the angle of detected Cherenkov
photons is translated into a Liklihood of a given photon being due
to a particular particle. Contributions from all photons
associated with a particular track are then summed to form an
overall Liklihood denoted as ${\cal L}_i$ for each particle
hypothesis. To differentiate between pion and kaon candidates, we
use the difference: $-2\log({\cal L_{\pi}})+2\log({\cal L}_K$).
Usually this cut is set at zero except for muon candidates where
the difference between $-2\log({\cal L_{\mu}})+2\log({\cal L}_K$)
is set at 10, to ensure a high, well understood efficiency. To
utilize the dE/dx information we calculate  $\sigma_{\pi}$ as the
difference between the expected ionization loss for a pion and the
measured loss divided by the measurement error.  Similarly,
$\sigma_{K}$ is defined  in the same manner using the expected
ionization for a  kaon .

We use both the RICH and dE/dx information for $D^-$ meson tag
candidate tracks in the following manner: (a) If neither the RICH
nor dE/dx information is available, then the track is accepted as both a pion and a kaon candidate. (b)
If dE/dx is available and RICH is not then we insist that pion
candidates have $PID_{dE}=\sigma_{\pi}^2-\sigma_{K}^2 <0$ and kaon
candidates have $PID_{dE}> 0.$ (c) If RICH information is available
and dE/dx is not available, then we require that
$PID_{RICH}=-2\log({\cal L}_{\pi})+2\log({\cal L}_K)<0$ for pions
and $PID_{RICH}>0$ for kaons. (d) If both dE/dx and RICH
information are available, we require that $(PID_{dE}+PID_{RICH})
<0$ for pions and $(PID_{dE}+PID_{RICH})>0$ for kaons.


We reconstruct $\pi^o$'s by first selecting photon candidates from
energy deposits in the crystals not matched to charged tracks that
have deposition patterns consistent with that expected for
photons. Pairs of photon candidates are kinematically fit to the
nominal $\pi^o$ mass.  We require the pull, the difference between
the raw and fit mass normalized by its uncertainty, to be less
than three for accepted $\pi^o$ candidates.

 $K_s^o$ are formed from a pair of charged pions  which are kinematically fitted to
 come from a single vertex. We also require that the invariant mass of the two
 pions be within 4.5 times the width of the $K_s$ mass peak which is
 0.004 GeV r.m.s.

\section{Reconstruction of Charged ${\boldmath D}$ Tagging Modes}

Tagging modes are fully reconstructed by first evaluating the
difference in the energy of the decay products with the beam
energy. We then normally require the absolute value of this
difference to be within 0.02 GeV of zero, approximately
twice the r.m.s. width, and then look at the
reconstructed $D^-$ beam constrained mass defined as
\begin{equation}
m_D=\sqrt{E_{beam}^2-(\sum_i\overrightarrow{p}_{\!i})^2},
\end{equation}
where $i$ runs over all the final state particles.
We also use the charge-conjugate $D^+$ tags and search for $D^-\to
\mu^-\overline{\nu}_{\mu}$; in the rest of this paper we will not
mention the charge-conjugate modes explicitly, but they are always
used.

\begin{figure}[htbp]
\centerline{ \epsfxsize=6.0in \epsffile{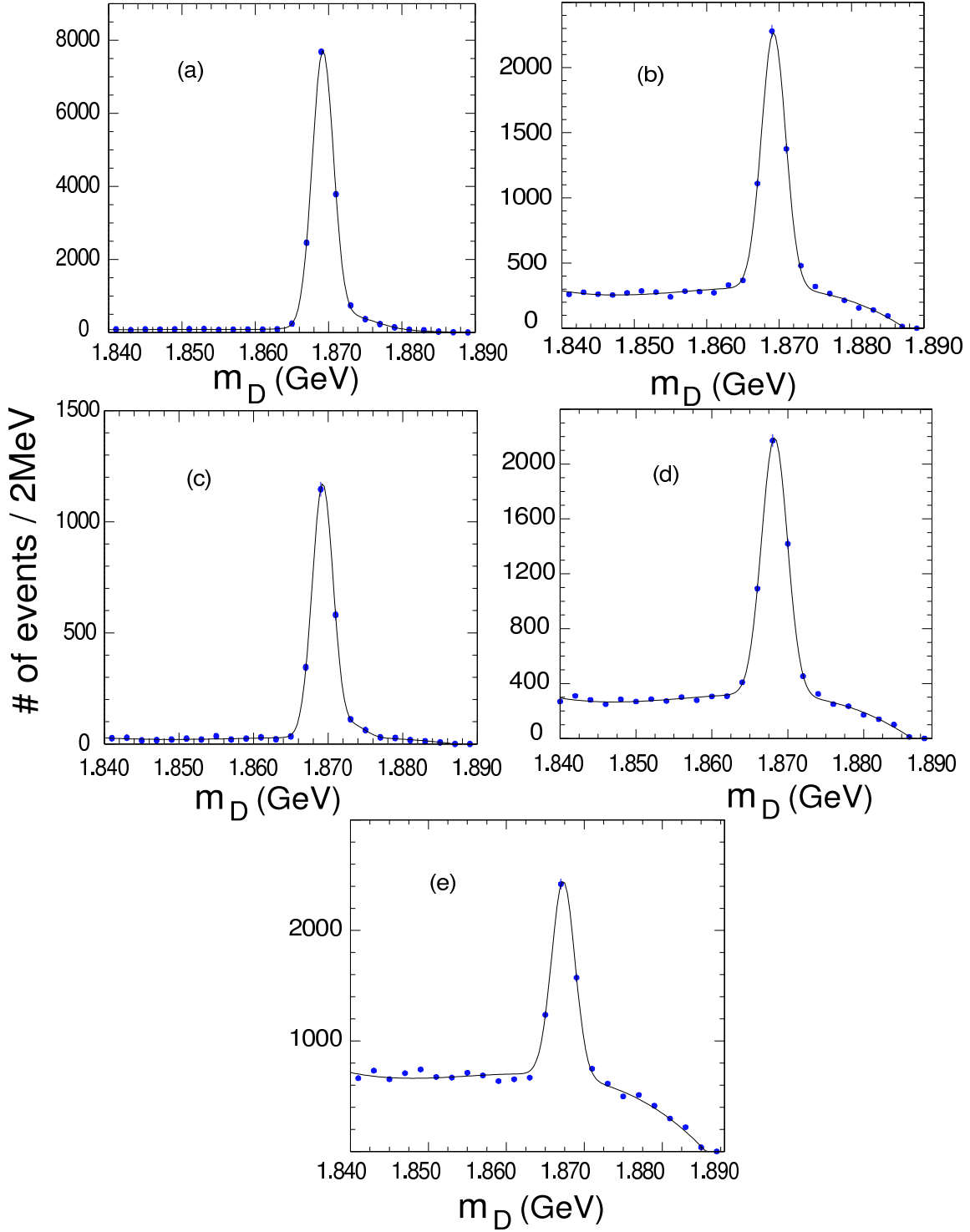} }
\caption{ Beam Constrained Mass distributions for different fully
reconstructed $D^-$ decay candidates; the curves show the sum of
Gaussian signal functions and $3^{rd}$ order polynomial background
functions. A single signal Gaussian is used for all modes except
for (a) and (c) where two Gaussians are used. (a) $D^- \to K^+
\pi^- \pi^-$, (b) $D^-\to K^+ \pi^- \pi^- \pi^0$, (c) $D^- \to K_s
\pi^-$, (d) $D^- \to K_s \pi^-\pi^-\pi^+$ and (e) $D^- \to
K_s\pi^- \pi^0$.(preliminary)} \label{Drecon}
\end{figure}

The $m_D$ distribution for all $D^-$ tagging modes considered in
this data sample are shown in Fig.~\ref{Drecon} and listed in
Table~\ref{tab:Drecon} along with the number of signal and
background events. The event numbers are determined from fits of
the $m_D$ distributions to Gaussian signal functions plus a
background shape parameterized as $3^{rd}$ order polynomial for
the $K^+\pi^-\pi^- \pi^o$, $K_s\pi^-\pi^-\pi^+ $ and
$K_s\pi^-\pi^o $ and from fits to double Gaussian signal functions
plus $3^{rd}$ order polynomial for $K^+\pi^-\pi^- $ and $K_s\pi^-$
tags where we see a small tail on the higher mass side.

\begin{table}[htb]
\begin{center}
\begin{tabular}{lcc}
    Mode  &  Signal           &  Background \\ \hline
$K^+\pi^-\pi^- $ & $15188 \pm 233$   & $583\pm 336$\\
$K^+\pi^-\pi^- \pi^o$ & $4082 \pm 81$  & $1826 \pm 343$\\
$K_s\pi^-$ &   $2110\pm 81$& $251\pm19$\\
$K_s\pi^-\pi^-\pi^+ $ &  $3975 \pm 81$ & $1880\pm 342$\\
$K_s\pi^-\pi^o $ &  $3297 \pm 87$ & $4226\pm 516$\\
\hline
Sum &  $ 28652\pm285$ & $8765\pm784$\\
\hline\hline
\end{tabular}
\end{center}
\caption{Tagging modes and numbers of signal and background events
determined from the fits shown in Fig.\ref{Drecon}.}
\label{tab:Drecon}
\end{table}

Selecting those candidates within 3 r.m.s. widths of the $D^-$
mass reduces the signal number by 77 events giving a total of
28575$\pm$286 events used for further analysis.



\section{${\boldmath D^+\to \mu^+\nu_{\mu}}$ Selection Criteria}
 To select $D^+\to \mu^+\nu_{\mu}$ events we first
reconstruct $D^-$ event candidates and then search for events with a single
additional charged track presumed to be a $\mu^+$. Then  we infer
the existence of the neutrino by requiring a measured value near
zero (the neutrino mass) of the missing mass squared (MM$^2$)
defined as
\begin{equation}
{\rm
MM}^2=\left(E_{beam}-E_{\mu^+}\right)^2-\left(-\overrightarrow{p}_{\!D^-}
-\overrightarrow{p}_{\!\mu^+}\right)^2, \label{eq:MMsq}
\end{equation}
where $\overrightarrow{p}_{D^-}$ is the three-momentum of the
fully reconstructed $D^-$.

We need to restrict the sample to candidate $\mu^+ \nu_{\mu}$
events resulting from the other $D$. Thus we wish to exclude
events with more than one additional charged track, which we take
to be the muon candidate, or with extra neutral energy. It is
possible, in fact even likely, that the decay products of the
tagging $D^-$ interact in the detector material, mostly the EM
calorimeter and spray tracks and neutral energy back into the rest
of the detector. To evaluate the size of these contributions we
use a very pure sample of events obtained by finding fully
reconstructed $D^o\overline{D}^o$ events. The numbers of these
events in various decay modes are listed in
Table~\ref{tab:double}, we use a total of 782 events.

\begin{table}[htb]
\begin{center}
\begin{tabular}{llc}
    Mode 1 &  Mode 2           &  \# of events \\\hline
$K^-\pi^+$  & $K^+\pi^-$  & 89 \\
$K^+\pi^-\pi^+\pi^-$  & $K^-\pi^+$  &392 \\
$K^+\pi^-\pi^+\pi^-$ & $K^-\pi^+\pi^-\pi^+$  & 301 \\
\hline\hline
\end{tabular}
\end{center}
\caption{Fully reconstructed $D^o\overline{D}^o$ events }
\label{tab:double}
\end{table}

The number of interactions of particles with material and their consequences depend
on the number of particles, the kind of particles and their momenta.
Thus, the sum over these neutral $D$ decay modes isn't quite the same as the
sum over the tagging $D^-$ decays, however the average over these modes is quite
 similar to the $D^-$ tagging modes for this level of
 statistics.

Extra tracks do appear in these $D^o\overline{D}^o$ events. None
of these tracks, however, approach the main event vertex.
Requiring that good tracks are within 0.05 m along the beam and
0.005 m perpendicular to the beam does not include any additional
tracks from interactions in the material. $D^-$ tags with
additional $K_s \to \pi^+\pi^-$ candidates are rejected.

\begin{figure}[htbp]
\centerline{ \epsfxsize=3.0in \epsffile{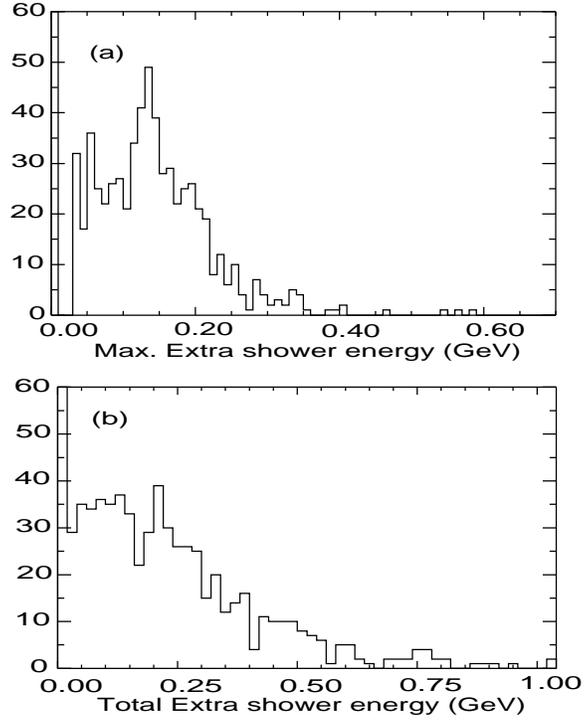} }
\caption{ Largest and Total Extra shower energies in the
$D^o\overline{D}^o$ sample.} \label{Esh}
\end{figure}

In the $D^o\overline{D}^o$ events, energy in the calorimeter not
matched to any of the charged tracks is shown in Fig.~\ref{Esh}.
Figure~\ref{Esh}(a) shows the energy of the largest shower and
~\ref{Esh}(b) shows the total.
 We accept only as extra showers those that do not match a charged
 track within a connected region.  A connected region  is a group of adjacent crystals
 with energy depositions which are nearest neighbors. This suppresses a lot of hadronic
 shower fragments which would otherwise show up as unmatched
 showers. Hadronic interactions and very
 energetic $\pi^o$'s tend to produce one connected region with many
 clusters.
 For further analysis we require that the largest unmatched shower
not to be larger than 250 MeV. This requirement is
(93.5$\pm$0.9)\% efficient for signal events.

\begin{figure}[htbp]
\centerline{ \epsfxsize=4.0in
\epsffile{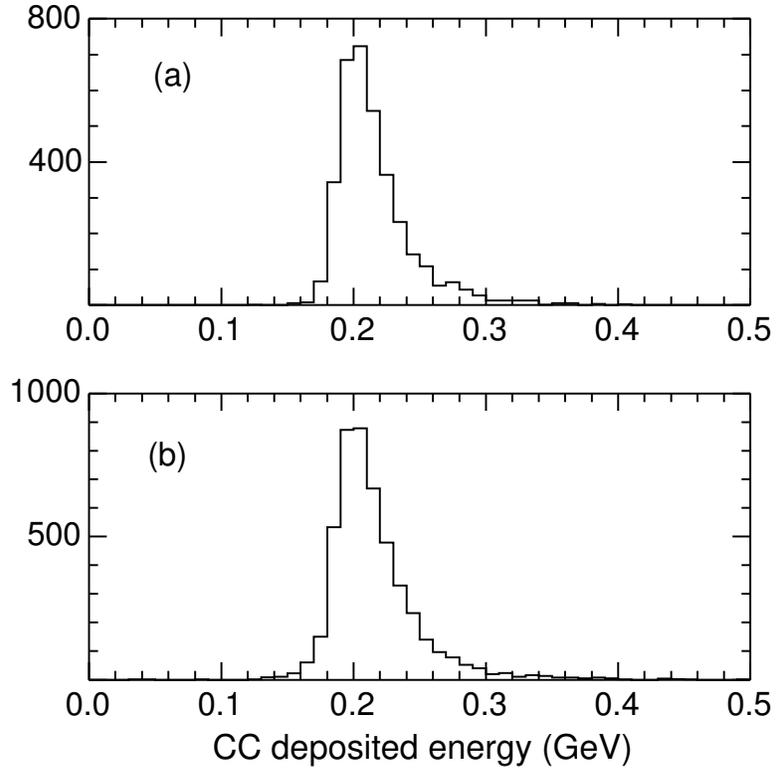} } \caption{ Muon
deposited energy in CC of Muon pairs from (a) DATA and (b) MC.}
\label{mu-data-mupair}
\end{figure}

The muon candidate is required to be within the barrel region of
the detector $|\cos\theta|<0.81,$ where $\theta$ is the angle of
the muon with respect to the beam electrons; this requirement
insures that the MM$^2$ resolution is good as tracks at larger
angles with respect to the beam are measured with poorer
precision.  In addition, this requirement helps reject background
from the decay $D^+\to\pi^+\pi^o$; this mode also gives a MM$^2$
near zero. Requiring the muon candidate in the barrel region (the
$\pi^+$ in this case) avoids having the photons from this decay
being lost in the transition region of the calorimeter between the
barrel and the endcap, because the $\pi^o$ direction is almost
directly opposite the $\pi^+$. Furthermore, the muon candidate is
required not to be consistent with the kaon hypothesis using RICH
information. Finally, we also require that the muon candidate
deposits less than 300 MeV of energy in the calorimeter,
characteristic of a minimum ionizing particle. This requirement is
very efficient for real muons, and rejects about 40\% of the pions
as determined using a sample of reconstructed $D^o\to K^-\pi^+$
decays. Fig.~\ref{mu-data-mupair} shows the muon deposited energy
in the EM calorimeter both from data on $e^+e^-\to \mu^+\mu^-$ and
from GEANT simulation of the same process. The Monte Carlo and
data are in excellent agreement. We therefore use a GEANT
simulation of $D^+\to\mu^+\nu$ with lower energy muons to
determine that the efficiency of the calorimeter energy cut is
$(98.7 \pm 0.19)\%$.
\begin{table}[htb]
\begin{center}
\begin{tabular}{lccccl}
Tag&MM$^2$&CC energy  & $-2\log({\cal L}_K)$ &$-2\log({\cal L}_{\mu}$) &$\mu^{\pm}$\\
&  (GeV$^2$) & of $\mu^+$(GeV)  & &  &\\
\hline
 $K\pi\pi\pi^o$     & ~0.032  &0.186 &-4.3   &-166.0  &~+\\
 $K_s\pi$& ~-0.019  &0.201 &0.00    &-140       &~-\\
 $K\pi\pi$      &-0.051   &0.190 &~31.9    &-252.9 &~+\\
 $K\pi\pi$     &-0.004   &0.221 &0.00   &-115.2   &~+\\
 $K_s\pi\pi^o$     &0.032   &0.164 &~-0.32   &~-130.6 &~-\\
 $K_s\pi\pi\pi$&~0.001 &0.245 &-11.7   &-138.9 &~+\\
 $K\pi\pi\pi^o$& ~0.002  &0.204 &-8.6    &-88.6  &~-\\
 $K_s\pi\pi^o$  & ~0.014  &0.208 &-8.3    &-113.0     &~+\\

\hline\hline
\end{tabular}
\end{center}
\caption{Muon Candidate Properties. }\label{tab:Muprop}
\end{table}

Table.~\ref{tab:Muprop} shows the properties of each muon
candidate from the 8 events in the signal region.
When evaluating MM$^2$ using equation~\ref{eq:MMsq}
there are two important considerations that are not obvious.
First of all, we explicitly need to take into account the crossing angle between the $e^+$ and $e^-$ beams.
This angle is about 4 mrad, varying slightly run to run; we use this information and Lorentz transform all
laboratory quantities to the center-of-mass. Secondly, we require that the reconstructed
$D^-$ have exactly the known $D^-$ mass.
This changes and improves somewhat our knowledge of the $D^-$ direction.

\begin{figure}[htbp]
\centerline{ \epsfxsize=5.0in \epsffile{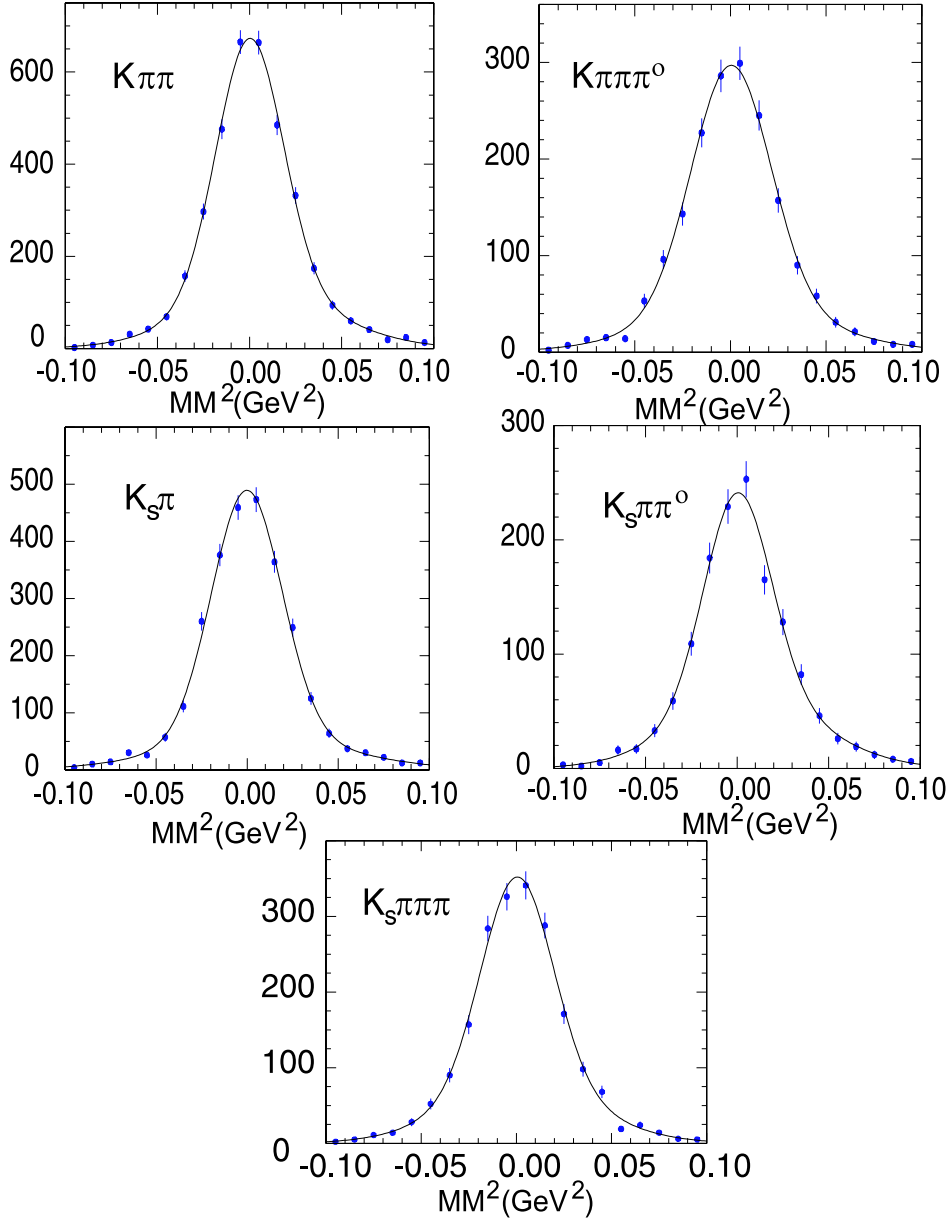} }
\caption{Monte Carlo simulation of $D^+\to \mu^+ \nu_{\mu}$ events for different
tags. The plots have been fitted to two Gaussians centered at
zero where the second Gaussian constitutes around 30\% of area.}
\label{mc-mm2}
\end{figure}

The MM$^2$ from Monte Carlo simulation is shown for our different tagging samples in Fig.~\ref{mc-mm2}. The signal is fit to a sum of two Gaussians with the wider Gaussian having about 30\% of the area independent of tagging mode. The resolution ($\sigma$) is defined as
\begin{equation}
\sigma = f_1\sigma_1+(1-f_1)\sigma_2,
\end{equation}
where $\sigma_1$ and $\sigma_2$ are the individual widths of the
two Gaussians and $f_1$ is the fractional area of the first
Gaussian. The resolution is approximately 0.025 GeV$^2$ consistent
among all the tagging decay modes.

\begin{figure}[htbp]
\centerline{ \epsfxsize=3.0in \epsffile{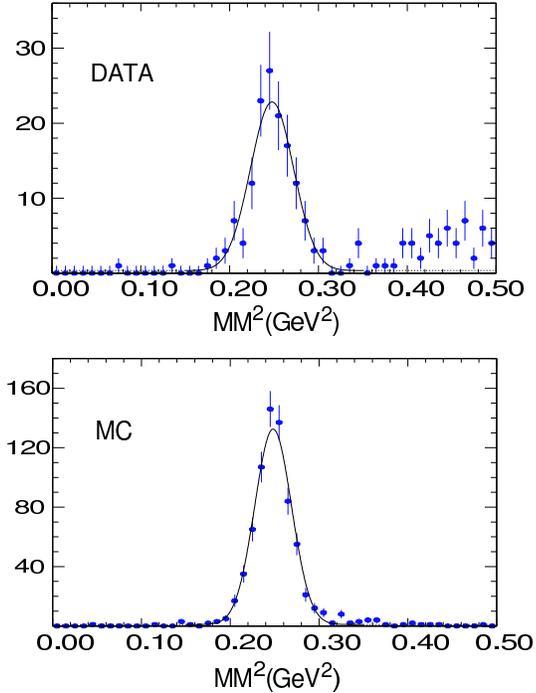} }
\caption{ MM$^2$ distribution for the decay $D^-\to K_s\pi^-$ from
data and MC} \label{mc-data-check}
\end{figure}

 We check our simulations using the $D^-\to K_s\pi^-$ decay. Here
 we choose events with the same requirements as used to search for
 $\mu^+\nu$ but require one additional found $K_s$. The MM$^2$
 distribution for this final state is shown in Fig.~\ref{mc-data-check}. The
 resolution is measured to be 0.024$\pm$0.002 GeV$^2$ from a
 single Gaussian fit, consistent with but slightly larger than the Monte
Carlo estimate of 0.021$\pm$0.001 GeV$^2$. To account for the
difference in resolution between data and simulations we scale the
resolution by 14 \% to 0.028 GeV$^2$ when looking for the
$D^+\to\mu^+\nu_{\mu}$ signal.

\begin{figure}[htbp]
\centerline{ \epsfxsize=3.0in \epsffile{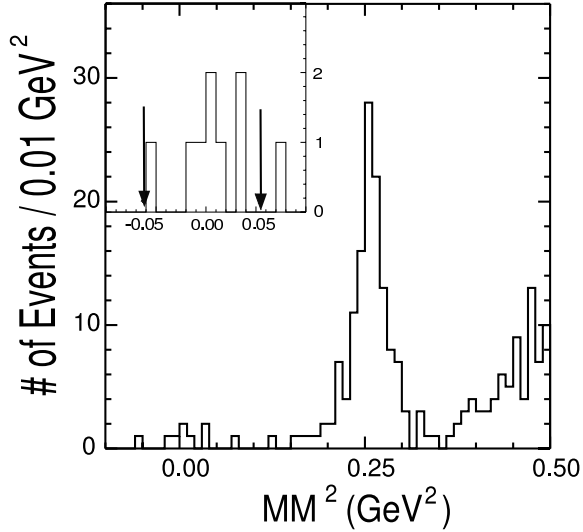} }
\caption{MM$^2$ using $D^-$ tags and one additional opposite sign
charged track and no extra energetic showers (see text). The
insert shows the signal region for $D^+\to\mu^+\nu$ enlarged, $\pm
2\sigma$ range is shown between the two arrows.(preliminary)}
\label{mm2}
\end{figure}

The MM$^2$ distributions for our tagged events requiring no extra
charged tracks besides the muon candidate and showers above 250
MeV as described above is shown in Fig.~\ref{mm2}.  We see a small
signal near zero containing 8 events within an interval that is
twice as wide as our resolution, -0.056 GeV$^2$ to +0.056 GeV$^2$.
This signal is most likely due to the $D^+\to\mu^+\nu_{\mu}$ mode
we are seeking. The large peak centered near 0.25 GeV$^2$ is from
the decay $D^+\to K^o\pi^+$ that is far from our signal region.

\begin{figure}[htbp]
\centerline{ \epsfxsize=4.0in
\epsffile{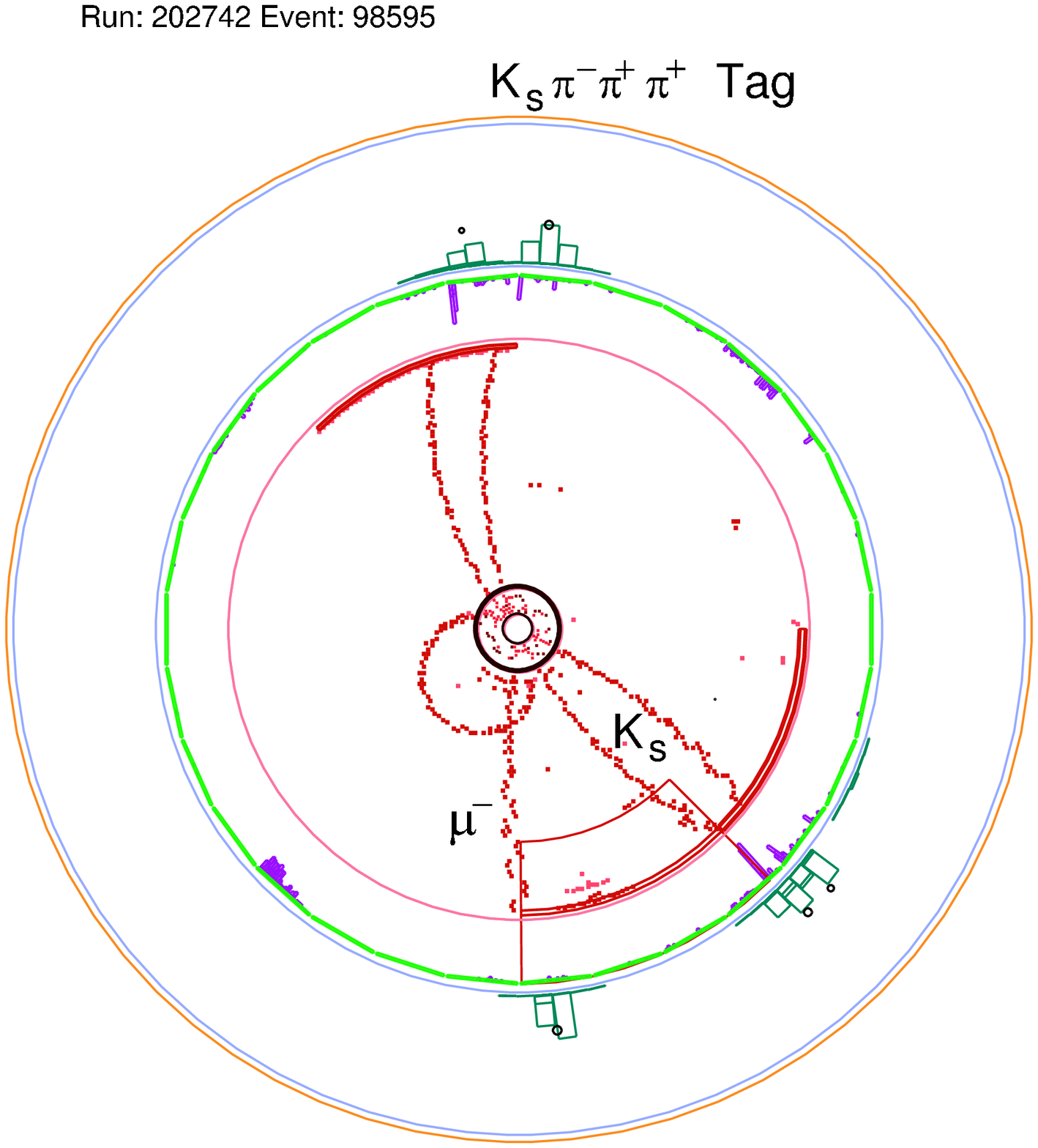} } \caption{ Typical
$D^-\to\mu^- \overline{\nu}_{\mu}$ event. The muon is shown as
well as the two charged pions forming the $K_s$. We notice the
presence of a curler track (a pion) with momentum around 50 MeV.}
\label{event}
\end{figure}

A typical event is shown in Fig.~\ref{event}.


\section{Background Evaluation}
\subsection{Introduction}

\begin{figure}[htbp]
\centerline{ \epsfxsize=4.0in \epsffile{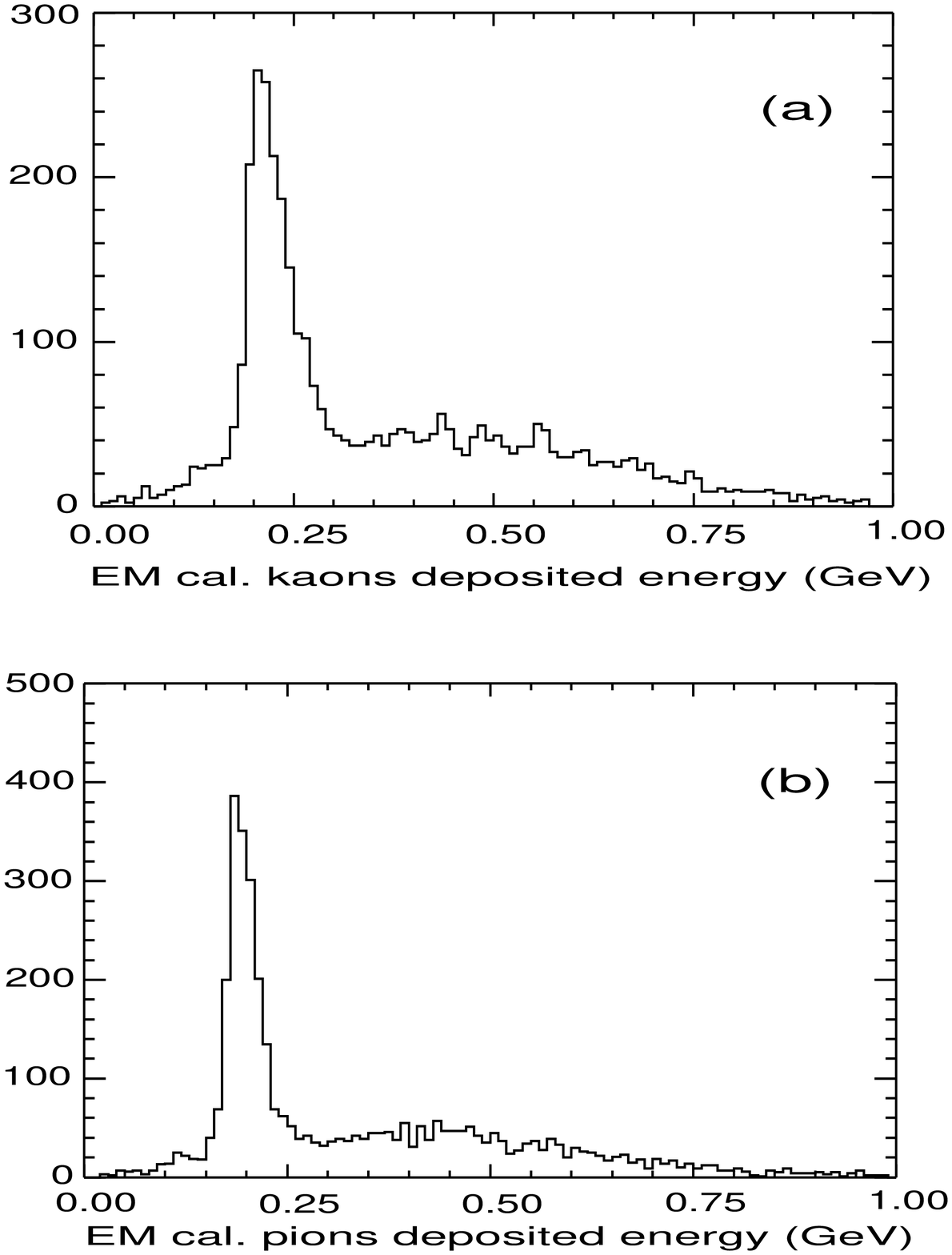} }
\caption{ Deposited energy in EM calorimeter for (a) Kaons, (b)
Pions from $D^0 \to K^-\pi^+$.} \label{kapi-dep}
\end{figure}

There are several background sources we need to evaluate. These
include background from other $D^+$ modes, background from
misidentified $D^o\overline{D}^o$ events and continuum background.
The requirement of the muon depositing $<$300 MeV in the
calorimeter, while about 99\% efficient on muons, rejects
only about 40\% of pions as determined from the
$D^o\overline{D}^o$ event sample where the pion from the
$K^{\pm}\pi^{\mp}$ mode was examined. In Fig.~\ref{kapi-dep} we
show the deposited energy in the calorimeter for both kaons and
pions obtained from the $K \pi$ tag sample.
\subsection{${\boldmath D^+}$ Backgrounds}

\begin{figure}[htbp]
\centerline{ \epsfxsize=4.0in \epsffile{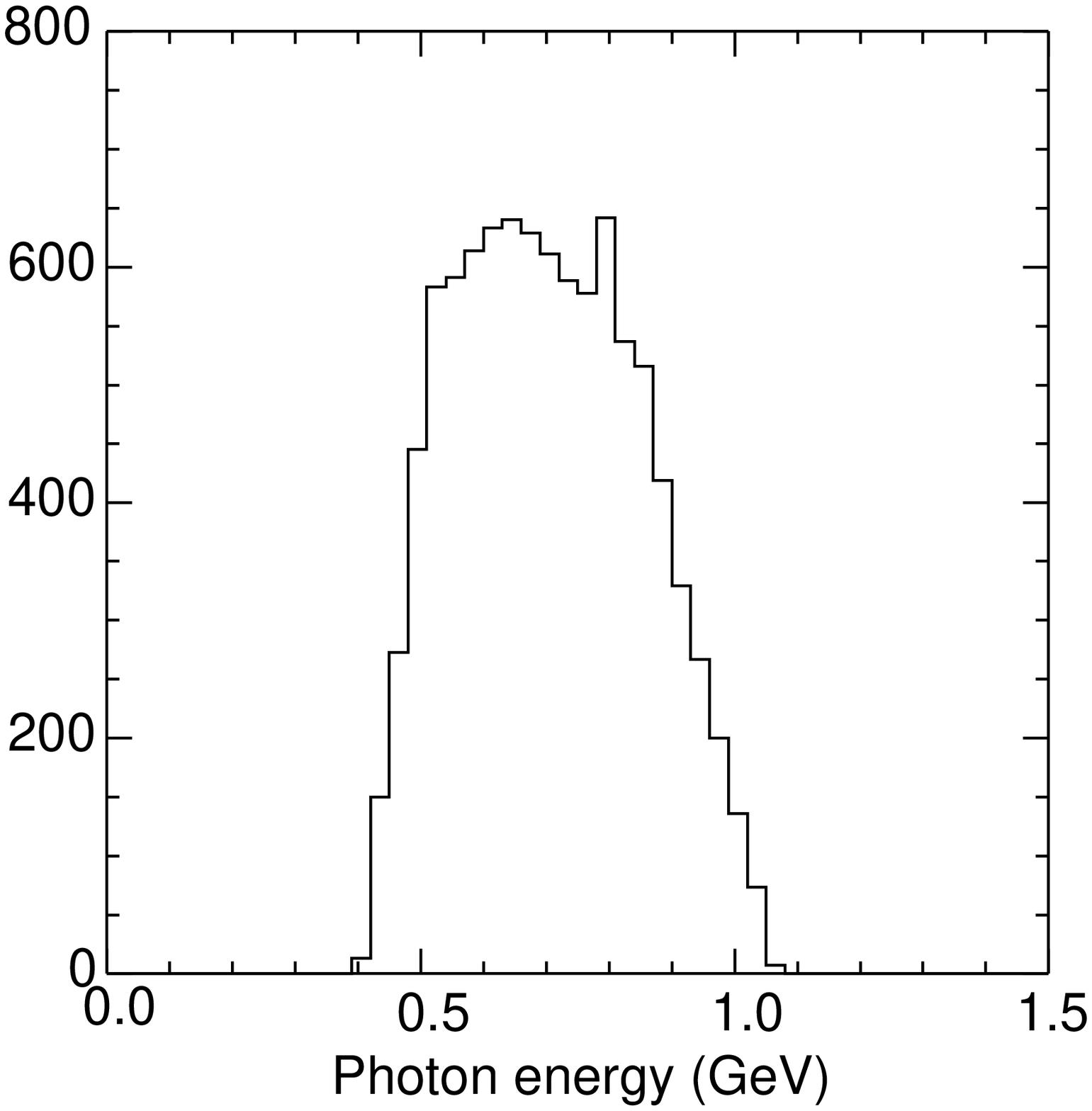} }
\caption{ Maximum photon energy of the $\pi^o$ in the $D^-\to
\pi^-\pi^o$ decay.} \label{mc-max}
\end{figure}

There are a few $D^+$ decay modes that could simulate the signal.
These are listed in Table~\ref{tab:Dpback} along with the
background estimate we obtained by Monte Carlo generation and
reconstruction of each specific mode. The branching ratios are
from the Particle Data Group except for the $\pi^+\pi^o$ mode
where a separate CLEO analysis gives a somewhat lower value
\cite{CLEOpipi}. This mode is the most difficult to reject because
the MM$^2$ peaks
 very close to zero, at 0.018 GeV$^2$, well within our resolution of 0.028 GeV$^2$. While we have insisted that
  the muon candidate be well within our acceptance, it is possible for the photons from the $\pi^o$ decay to
   inadvertently be matched to the tracks from the tagging $D^-$ or be
   missed. The maximum photon energy of the $\pi^o$ at the generator level is shown in
   Fig.~\ref{mc-max}. We note that at least one photon from the
   $\pi^+\pi^o$ mode exceeds our 250 MeV calorimeter
   energy requirement and should in most cases cause such a decay
   to be vetoed.

\begin{figure}[htb]
\centerline{
\epsfxsize=4.0in \epsffile{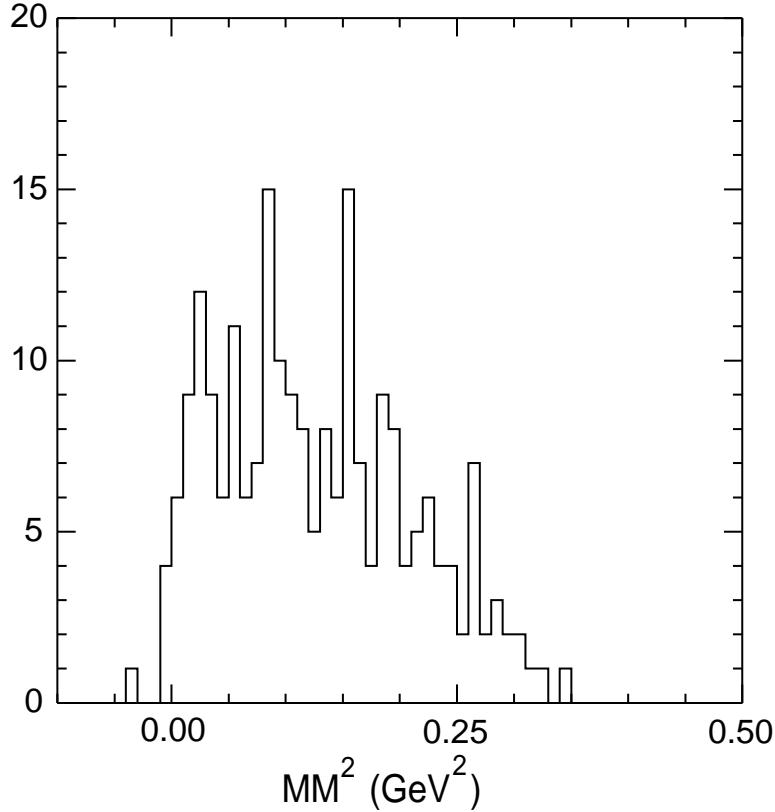}}
\caption{Missing
Mass squared distribution for $D^+ \to \tau^+\nu$ and $\tau^+\to
\pi^+\nu$.} \label{tau}
\end{figure}

   Even though the $K^0\pi^+$ mode gives a large peak in the
   MM$^2$spectrum near 0.025 GeV$^2$, our simulation shows that
   only a very small amount can enter our signal region, only 0.06
   events.
We have simulated backgrounds from $D^+\to\tau^+\nu$. Out of 10,000 simulated
events with $D^-$ tags, we found background only when $\tau^+\to \pi^+\nu$.
Because of
the small $D^+$-$\tau^+$ mass difference, the $\tau^+$ is almost at rest in the
laboratory frame and thus the $\pi^+$ has relatively large momentum causing the
MM$^2$ distribution to populate only the low MM$^2$ region, even in this case
with two missing neutrinos. The MM$^2$ distribution is
shown in Fig.~\ref{tau}.

\begin{table}[htb]
\begin{center}
\begin{tabular}{lll}
    Mode & ${\cal{B}}$ (\%) & \# of events \\\hline
$\pi^+\pi^o $ &0.13$\pm$0.02   & 0.31$\pm$0.04 \\
$K^o\pi^+ $&  2.77$\pm$0.18  &0.06$\pm$0.05\\
$\tau^+\nu$& 3.2$\times$ $\mu^+\nu$ & 0.36$\pm$0.08 \\
$\pi^o\mu^+\nu$& $0.31\pm 0.15$  & negligible \\
\hline\hline
\end{tabular}
\end{center}
\caption{Backgrounds from specific $D^+$ decay modes }
\label{tab:Dpback}
\end{table}

The semileptonic mode $\pi^o\mu^+\nu_{\mu}$ is similar to
$\pi^+\pi^o$ except that the $\pi^o$ often carries off enough
momentum to result in large MM$^2$. We found no candidate
background events in a Monte Carlo sample consisting of 50,000
tags plus a $D^+\to \pi^o\mu^+\nu$ decay.

\subsection{{$\boldmath D^o\overline{D}^o$} and Continuum Backgrounds}

These backgrounds are evaluated by analyzing Monte Carlo samples
corresponding to 15 times the total amount of data in our
possession. To normalize our Monte Carlo events to our data sample
we used $\sigma_{D^o\overline{D}^o}=2.9$ nb and
$\sigma_{continuum}=6.4$ nb. In each sample we found one
background event within two standard deviations of zero. These
correspond to 0.16$\pm$0.16 $D^o\overline{D}^o$ events and
0.17$\pm$0.17 continuum events forming background.

\subsection{Background Summary}

Our total background is 1.07$\pm$0.25 events. Because of
the uncertainties in the Monte Carlo simulation we assign a 100\%
error to our background estimate: 1.07$\pm$1.07 events.

\section{Branching Ratio and Decay Constant}

Subtracting the 1.07 event background from our 8 events in the
signal region, we determine a branching fraction using a detection
efficiency for the single muon of 69.9\%. This efficiency includes
the tracking, the particle identification, the probability of the
crystal energy being less than 300 MeV and the probability of not
having another unmatched shower in the event with energy greater
than 250 MeV. We assign a relative 5.3\% error on this efficiency,
the components of which are shown in Table~\ref{tab:eff}. We use a
$3\%$ systematic error on track finding found using the double
tagged events and we estimated the error on the particle
identification cut to be $1\%$ from studies of $D^{*+}$ decays in
higher beam energy data. The error on the minimum ionization cut
is evaluated from the efficiency of the cut by generating signal
Monte Carlo. To evaluate the error on the extra shower cut we
evaluated the efficiency of that cut in the $D^o\overline{D}^o$
sample discussed above. The total systematic error is obtained by
summing all entries in quadrature.
\begin{table}[htb]
\begin{center}
\begin{tabular}{lc}
     & Systematic error \% \\ \hline
MC statistics &0.8  \\
Track finding &3 \\
PID cut &1 \\
Minimum ionization cut &1 \\
Extra showers cut &4  \\
Number of tags &1\\
 \hline\hline
\end{tabular}
\end{center}
\caption{Systematic errors on $D^+ \to \mu^+ \nu_{\mu}$
efficiency.} \label{tab:eff}
\end{table}

To compute the branching ratio we use 6.93 signal events divided
by 69.9\% and the 28575$\pm$286 $D^{\mp}$ tags. No other
efficiencies enter. We find
\begin{equation}
{\cal{B}}(D^+\to\mu^+\nu_{\mu})=(3.5\pm 1.4 \pm 0.6)\times
10^{-4}~,
\end{equation}
 The error on the background contributes 15.4\% to the systematic
 error on the branching ratio.

The decay constant $f_{D^+}$ is then obtained from Equation~\ref{eq:equ_rate} as
\begin{equation}
f_{D^+}=(201\pm 41\pm 17)~{\rm MeV}~.
\end{equation}

\section{Conclusions}
There have been several experimental studies of $D$ meson decay
constants. The Mark III group published an upper limit of
${\cal{B}}(D^+\to\mu^+\nu_{\mu})<7.2 \times 10^{-4}$, which leads to an upper limit on the
decay constant  $f_{D^+}<290$ MeV at 90\% confidence level based
on 9.3 pb$^{-1}$ of data taken on the $\psi''$ \cite{MarkIII}. BES
claimed the observation of one event at a center-of-mass energy of
4.03 GeV with a branching ratio of
$(0.08^{+0.17}_{-0.05})$\% \cite{Rong1}. Recently, using
17.7 pb$^{-1}$ of $\psi''$ data they presented a 3 event signal
and claimed a background of 0.25 events where neither $\pi^+\pi^o$, or $\tau^+\nu$ were
mentioned as a possible background mode, nor was continuum background
considered \cite{Rong2}. Here they find a branching ratio of
$(0.12^{+0.092+0.010}_{-0.063-0.009})$\%, and a
corresponding value of $f_{D^+}=(365^{+121+32}_{-113-28})$ MeV.
Our value is considerably smaller, though compatible with their
large error.

Our analysis shows the first statistically compelling signal for
$D^+\to\mu^+\nu$. The preliminary branching fraction is
\begin{equation}
{\cal{B}}(D^+\to\mu^+\nu)=(3.5\pm 1.4 \pm 0.6)\times 10^{-4}~,
\end{equation}
and the preliminary decay constant is
\begin{equation}
f_{D^+}=(201\pm 41\pm 17)~{\rm MeV}~.
\end{equation}

Our result for $f_{D^+}$, at the current level of precision, is consistent with predictions
of all of the models listed in Table~\ref{tab:Models}.

\begin{table}[htb]
\begin{center}
\begin{tabular}{lcc}
    Model  &  $f_{D^+}$ (MeV)          &  $f_{D_s^+}/f_{D^+}$           \\\hline
Lattice (unquenched) (UKQCD) \cite{Lattice:UKQCD} & $210\pm 10^{+17}_{-16}$ & $1.13\pm 0.02^{+0.04}_{-0.02}$\\
Lattice (partially quenched) (MILC) \cite{Lattice:Milc} &
$215\pm 6^{+16+8+4}_{-15-3-0}$&$1.14\pm 0.01^{+0.02}_{-0.03}\pm 0.03 \pm 0.01$ \\
QCD Spectral Sum Rules \cite{Chiral} & $203\pm 20$ & $1.15\pm 0.04$ \\
QCD Sum Rules \cite{Sumrules} & $195\pm 20$ & \\
Relativistic Quark Model \cite{Quarkmodel} & $243\pm 25$ & 1.10 \\
Potential Model \cite{Equations} & 238  & 1.01 \\
Isospin Mass Splittings \cite{Isospin} & $262\pm 29$ & \\
\hline\hline
\end{tabular}
\end{center}
\caption{Theoretical predictions of $f_{D^+}$ and
$f_{D_s^+}/f_{D^+}$} \label{tab:Models}
\end{table}

The models generally predict $f_{D_s^+}$ to be 10-15\% larger than
$f_{D^+}$. CLEO previously measured $f_{D_s^+}$ as ($280\pm 19\pm
28\pm 34)$ MeV \cite{chadha}, and we are consistent with these
predictions as well. We look forward to more data to improve the
precision.

\end{document}